\documentclass[twocolumn,showpacs,preprintnumbers,amsmath,amssymb]{revtex4}
\usepackage{graphicx}
\usepackage{amsmath}

\begin{document}

\title{ \bf Free spin quantum computation with semiconductor nanostructures}

\author{\bf  Wei-Min Zhang$^{a,b}$\footnote{E-mail: wzhang@mail.ncku.edu.tw},
Yin-Zhong Wu$^{a,c}$ and Chopin Soo$^{a,b}$}

\affiliation{$^{a}$Department of Physics and Center for Quantum
Information Science, \\ National Cheng Kung University, Tainan
70101, Taiwan \\
$^b$National Center for Theoretical Science, Tainan 70101, Taiwan
\\
$^{c}$Department of Physics, Changshu Institute of Technology,
Changshu 215500, P.~R.~China }

\begin{abstract}

Taking the excess electron spin in a unit cell of semiconductor
multiple quantum-dot structure as a qubit, we can implement
scalable quantum computation without resorting to spin-spin
interactions. The technique of single electron tunnelings and the
structure of quantum-dot cellular automata (QCA) are used to
create a charge entangled state of two electrons which is then
converted into spin entanglement states by using single spin
rotations. Deterministic two-qubit quantum gates can also be
manipulated using only single spin rotations with help of QCA. A
single-short read-out of spin states can be realized by coupling
the unit cell to a quantum point contact.
\end{abstract}

\pacs{73.63.-b, 03.67.Mn, 03.67.Lx}

\maketitle

The idea of using electron spins in semiconductor quantum dots as
qubits\cite{PRA1998} has received tremendous attention in the
implementation of scalable quantum computation. Recent experiments
showing unusually long spin decoherence time in
semiconductors\cite{book2,GaAs} provide a strong support for
pursuing this idea. Up to date, several quantum computation
schemes based on electron spins have be proposed with tunable
Heisenberg type spin-spin interactions in semiconductor
nanostrcutures\cite{nature1,Sham02,PRL,PRL1}. However, achieving a
tunable spin-spin interaction with sufficient strengthes
(comparing to the Coulomb interaction) is technically difficult.
An interaction free mechanism for logical operations on electron
spins is therefore more desirable. A few years ago, Knill,
Laflamme and Milburn show in an influential paper \cite{milburn}
that quantum computation can be implemented with photons using
only linear optics operators and single-photon detectors with
feedback. The situation is quite different for free fermions
according to the no-go theorem\cite{no-go1,no-go2}. Very recently,
Beenakker et al.~\cite{free} show that for free flying fermions,
one is able to construct a CNOT (controlled NOT) gate using only
spin beam splitters and single spin rotations if charge detectors
are added.

In this letter, we shall propose an implementation of scalable
spin quantum computation without resorting to spin-spin
interactions. We use external electrodes to control single
electron tunnelings that create naturally a charge entangled state
of two electrons with the help of a multiple-quantum-dot
structure, the quantum-dot cellular automata (QCA). The charge
entangled state is then converted into a spin entangled state of
the electrons using only single spin rotations (we call it as
charge-to-spin convention of electron entanglement states).
Spin-spin interactions are not required in this implementation and
deterministic two-qubit controlled gates can be easily
manipulated. Thus, a free spin quantum computation is feasible
with semiconductor nanostructurs.
\begin{figure}[ht]
   \centering
   \includegraphics[width=3.0in]{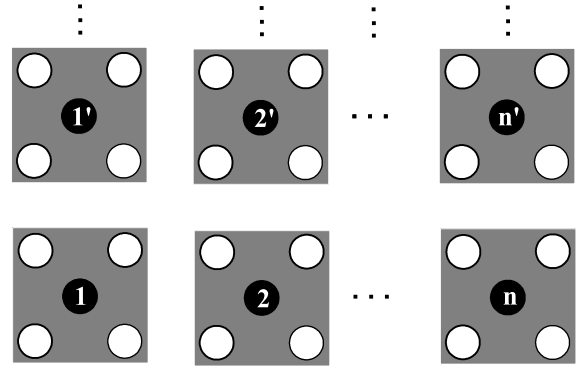}
   \caption{Schematic of a scalable free spin quantum computer
based on semiconductor nanostructures. Each shading square box
(contains five quantum dots) is taken as a unit cell.}
 \end{figure}

The architecture of our scalable quantum computer is based on the
semiconductor multiple quantum dot structures schematically shown
in Fig.~1. The basic devices (the shading boxes in Fig.~1) are
considered as unit cells. Each cell contains a qubit quantum dot
(the central black dot) surrounding with four ancilla dots (the
empty dots). The detailed structure of the cell $i$ is given by
Fig.~2a. The lines between quantum dots in a cell indicate the
possibility of interdot tunneling. The energy barriers between the
neighboring cells must be high enough to ensure that tunnelings of
electrons between different cells are forbidden. We also assume
that each cell is charged with only one excess conductor electron
and each dot in the cell is considered as a site. The on-site
charge energy $E_0$ of the excess electron in the qubit dot is low
enough comparing with the on-site energy $E_a$ in the ancilla dots
($\varepsilon=E_a-E_0>0$) such that the excess electron will sit
initially in the qubit dot due to the Coulomb blockade effect (see
Fig.~2b). Furthermore, the four ancilla dots within the cell are
coupled to the qubit dot through bias electrodes such that by
tuning on the bias voltage $V_i^{LR}$ or $V_i^{TB}$ (the anti-bias
voltage is given by $-V_i^{LR}$ or $-V_i^{TB}$), the excess
electron will tunnel coherently into the right or the bottom (the
top or the left for an anti-bias voltage) two ancilla dots with
equal possibility if the ancilla dots are fabricated identically
(also see Fig.~2c). This architecture can be achieved by
current/further development of nano-technology.
\begin{figure}[ht]
   \centering
   \includegraphics[width=3.3in]{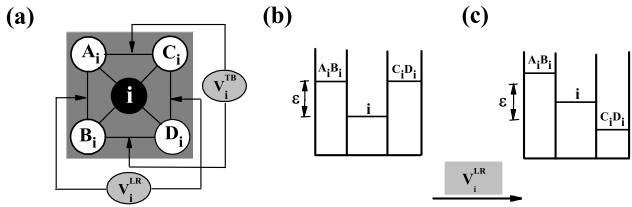}
   \caption{(a) The detailed structure of the unit cell $i$.
The bias electrode voltages $V_i^{LR}$ and $V_i^{TB}$ acting on
the cell control the electron tunnelings among dots within the
cell; (b) and (c) show the on-site energy of the electron in
different quantum dots in the cell without and with the bias
voltage $V_i^{LR}$ respectively.}
 \end{figure}

Based on such an architecture of the basic quantum devices, the
four empty dots between two qubit dots of the neighboring cells
(given by the dotted square box, e.g.~C$_i$-D$_i$-B$_j$-A$_j$ in
Fig.~3) form a usual structure of QCA \cite{lent93}. QCA has been
used to simulate classical digital algorithms \cite{science2}. A
semiconductor realization of such a structure has also been
developed\cite{SQCA}. Quantum mechanically, when a QCA is charged
with two electrons, these two electrons will occupy coherently two
diagonal sites (two charge polarizations) as a result of Coulomb
repulsion\cite{QCQCA}. The corresponding two-electron charge state
can be in an arbitrary superposition state of the two charge
polarizations ($P=\pm 1$) which creates indeed an entangled charge
state of the two electrons. In additional, an external bias
polarization $E_{\rm bias}$ is coupled to each QCA for adjusting
the splitting of two charge polarizations\cite{QCQCA}, as shown in
Fig.~3.
\begin{figure}[ht]
   \centering
   \includegraphics[width=3.3in]{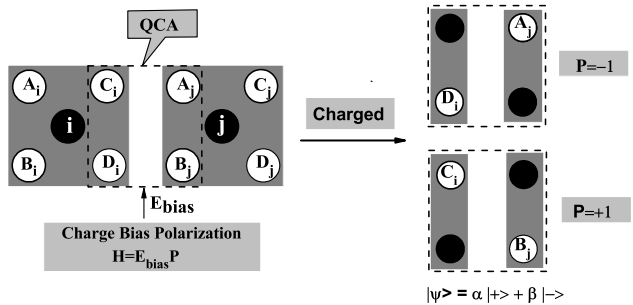}
   \caption{Quantum mechanically, the four quantum dots (the dotted
square boxes) between two qubit dots of neighboring cells form a
coherent QCA.}
 \end{figure}

We shall use this quantum mechanical QCA structure to manipulate
electron spin entangled states and two-qubit control gates through
single electron tunnelings and single spin rotations only. To be
explicit, we define quantum states of the excess electron in each
cell as a direct product of electron spin and charge states $|
S_{i} \rangle |e_{i}\rangle $ ($i=1,2...$), where the electron
sits initially in the qubit dot. The spin states of the excess
electron $| S_{i}\rangle $ are chosen as the qubit states in Pauli
basis $|\uparrow \rangle=|0 \rangle$ and $|\downarrow \rangle =|1
\rangle$, the electron charge state $|e_i \rangle$ is considered
as an ancilla state. A static uniform magnetic field is applied to
split the qubit states $|0 \rangle$ and $|1 \rangle$ by Zeeman
energy. The excess electron in each cell can be driven away from
its initial site (the qubit dot) in ancilla dots only when a
two-qubit controlled operation is performed, and will be pushed
back to the initial site as soon as the two-qubit operation is
completed, this is controlled by the bias voltage pulses
$V_i^{LR}, V_i^{TB}$. To distinguish the different sites of the
electron in the cell, we denote the charge state of the electron
siting in the ancilla dots A$_i, B_i, C_i$, and $D_i$ by
$|e_i^X\rangle$ with $X=A, B, C$ and $D$, respectively. The site
dependence of electron spin state is ignorable.

Now, two-qubit controlled operations can be implemented as
follows: consider a pair of neighboring unit cells, e.g.~the
$i$-th and the $j$-th cells in Fig.~3. The initial state of the
two excess electrons is given by
\begin{equation} \label{initialstate}
|\Psi_0 \rangle = |S_ iS_j \rangle |e_i e_j \rangle ,
\end{equation}
By tuning on the bias voltage $V_i^{LR}$ and the anti-bias voltage
$-V_j^{LR}$ to lower the electron potential energy of the dots
$C_i, D_i$ and $A_j, B_j$, the excess electron in each cell is
tunneled with definite probabilities into the quantum dots $C_i,
D_i$ and $A_j, B_j$, respectively. If we assume that the ancilla
dots in each cell are identical, the tunneling rates of the excess
electron into the dots $C_i$ and $D_i$ ($A_j$ and $B_j$) in the
cell $i$ ($j$) are equal. As mentioned before, the four quantum
dots ($C_i, D_i, A_j, B_j$) form quantum mechanically a coherent
QCA. When the QCA is charged with two excess electrons, these two
electrons will occupy coherently two diagonal sites as a result of
Coulomb repulsion. Since the electron tunnelings between different
cells are forbidden, the two electron charge state becomes a
superposition state of the two charge polarization states,
\begin{equation} \label{chargee}
|\Psi_0 \rangle \stackrel{(V_i^{LR},-V_j^{LR})_{\rm
on}}{\longrightarrow} | \Psi_1 \rangle=|S_i S_j \rangle
\frac{1}{\sqrt{2}} \big(|e_i^C e_j^B \rangle + |e_i^D e_j^A
\rangle \big) .
\end{equation}
It shows that with the help of QCA, the two electron charge state
becomes a maximally entangled state after switching on the bias
voltages $V_i^{LR}$ and $-V_j^{LR}$.

Now, we can convert the charge entangled state into a spin
entangled state using only single spin rotations. Explicitly,
consider the initial spin state of the two electrons: $|S_i S_j
\rangle=|\uparrow \downarrow \rangle =|0 1\rangle$. We take a spin
rotation $R_x(\pi)$ on each electron sited at the dots $D_i$ and
$A_j$ respectively after Eq.~(\ref{chargee}), where
$R_x(\theta)\equiv \exp(-i\theta\sigma_x/2)$. The corresponding
spin state of the two electrons becomes $R_x^D(\pi)R_x^A(\pi)|01
\rangle= - |1 0 \rangle$. Following the rotation operations, we
push the two electrons back into the qubit dots $i$ and $j$ by
switching off the bias voltages $V_i^{LR}$ and $-V_j^{LR}$. Thus,
the charge states return back to the initial states: $|e_i^C e_j^B
\rangle\rightarrow |e_i e_j \rangle$ and $|e_i^D e_j^A
\rangle\rightarrow |e_i e_j \rangle$. Consequently, we have
\begin{eqnarray} \label{bell1}
|\Psi_1 \rangle  & \stackrel{R_x^D(\pi)R_x^A(\pi)}
{\longrightarrow} & \frac{1}{\sqrt{2}} \big(|0 1 \rangle |e_i^C
e_j^B \rangle -|10 \rangle|e_i^D e_j^A \rangle \big) \nonumber \\
& \stackrel{(V_i^{LR},-V_j^{LR})_{\rm off}}{\longrightarrow}&
 \frac{1}{\sqrt{2}}\big(|0 1 \rangle  - |10 \rangle \big)|e_i e_j
 \rangle ,
\end{eqnarray}
namely, the electron charge entangled state has been converted
into a spin Bell state $|\psi^-\rangle$. Here, tuning on and
tuning off the bias voltages $V_i^{LR}$ and $-V_j^{LR}$ can be
achieved by bias electrode pulses\cite{Fujikawa}, the single spin
rotations can be implemented by either local magnetic fields or
ultrafast optical pulses as we shall discuss later. The pulse
sequence for generating the above spin Bell state through QCA is
illustrated in Fig.~4a. Repeating the process of
Eqs.~(\ref{initialstate}-\ref{bell1}) with different initial spin
states and single spin rotations, we can generate other three spin
Bell states ($|\psi^+\rangle, |\phi^-\rangle, |\phi^+\rangle$):
\begin{eqnarray} \label{bell2}
|0 1 \rangle |e_i e_j
\rangle  & \stackrel{R_x^D(\pi)R_x^A(3\pi)} {\longrightarrow} &
\frac{1}{\sqrt{2}}\big(|0 1 \rangle  + |1 0
\rangle\big)|e_i e_j \rangle, \nonumber \\
|0 0 \rangle |e_i e_j \rangle  & \stackrel{R_x^D(\pi)R_x^A(\pi)}
{\longrightarrow} & \frac{1}{\sqrt{2}}\big(|0 0 \rangle  - |1 1
\rangle\big)|e_i e_j \rangle , \\
|0 0 \rangle |e_i e_j \rangle  & \underbrace{
\stackrel{R_x^D(\pi)R_x^A(3\pi)}
{\longrightarrow}}_{V_i^{LR},-V_j^{LR}} &
\frac{1}{\sqrt{2}}\big(|0 0 \rangle  + |1 1 \rangle\big)|e_i e_j
\rangle. \nonumber
\end{eqnarray}
\begin{figure}[ht]
   \centering
   \includegraphics[width=2.6in]{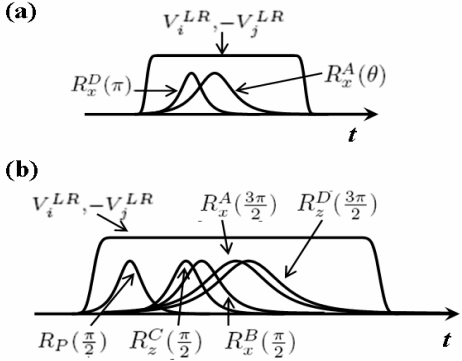}
   \caption{(a) An illustration of the pulse process for creating spin Bell
states through QCA, where $\theta=\pi$ or $3\pi$; (b) The pulse
process for CNOT gate.}
 \end{figure}

To construct a two-qubit controlled gate, we shall modify first
the charge entangled state in Eq.~(\ref{chargee}) with the help of
the bias polarization $E_{\rm bias}$. The charge polarizations of
two-electron states in QCA are defined as $|e_i^C e_j^B
\rangle=|-\rangle$ and $|e_i^De_j^A \rangle = | +\rangle$ with
polarizations $P=\mp 1$. The bias polarization $E_{\rm bias}$
coupling to QCA results in an effective Hamiltonian $H_P=E_{\rm
bias}\hat{P}$ \cite{QCQCA}, where $\hat{P} |+ \rangle = |+
\rangle$ and $\hat{P} |-\rangle = -|- \rangle$. Thus, applying a
bias polarization $\pi/2$-pulse to Eq.~(\ref{chargee}), we obtain
\begin{equation}  \label{poli}
|\Psi_1 \rangle \stackrel{R_P({\pi \over 2})}{\longrightarrow}
e^{i\pi/4}|S_i S_j \rangle \frac{1}{\sqrt{2}} \big( |e_i^C e_j^B
\rangle -i |e_i^De_j^A \rangle \big).
\end{equation}
Then, we apply further a single spin rotation on each dot $C_i$,
$D_i$, $A_j$, $B_j$ in the QCA with the operators
$R_z^C(\frac{\pi}{2})$, $R_z^D(\frac{3\pi}{2})$,
$R_x^A(\frac{3\pi}{2})$, $R_x^B(\frac{\pi}{2})$ respectively to
rotate the corresponding electron spin states. Finally, by
switching off the bias voltages $V_i^{LR}$ and $-V_j^{LR}$, the
two excess electrons return back to the qubit dots in the cells
$i$ and $j$. With these operations, the initial state
Eq.~(\ref{initialstate}) becomes
\begin{equation}
 \begin{array}{c}
|0 0 \rangle |e_i e_j \rangle \\ ~\\
|0 1 \rangle |e_i e_j \rangle \\ ~\\
|1 0 \rangle |e_i e_j \rangle \\ ~\\
|1 1 \rangle |e_i e_j \rangle \\
\end{array}
~~\underbrace{ \stackrel{U_P U_S(ABCD)}
{\longrightarrow}}_{V_i^{LR},-V_j^{LR}}~~
\begin{array}{c}
|0 0 \rangle |e_i e_j \rangle \\ ~\\
|0 1 \rangle |e_i e_j \rangle \\ ~\\
|1 1 \rangle |e_i e_j \rangle \\ ~\\
|1 0 \rangle |e_i e_j \rangle \\
\end{array}
\end{equation}
where, $U_P \equiv R_P({\pi \over 2}), U_S(ABCD) \equiv
R_x^A(\frac{3\pi}{2})\otimes R_x^B(\frac{\pi}{2})\otimes
R_z^C(\frac{\pi}{2})\otimes R_z^D(\frac{3\pi}{2})$. Thus, a
two-spin qubit CNOT gate is manipulated using only single spin
rotations through QCA for single electron tunnelings. The pulse
sequence for such an manipulation is given in Fig.~4b.

The above discussion shows that the spin two-qubit controlled
operations can be achieved using only single spin rotations and
single electron tunnelings through QCA. The method of using bias
electrodes to control single electron tunnelings has been
investigated in recent years for single electron transistor (SET)
devices \cite{SET} and also for the manipulation of single
electron charge qubit in double quantum dots\cite{Fujikawa} at a
time scale of a few hundred picosecond. Using bias electrodes to
control electron charge polarizations in QCA  has also been
proposed and discussed\cite{QCQCA}. The single spin rotations in
semiconductor quantum dots have been extensively
explored\cite{book2}. The simplest manipulation of a single spin
is to expose individual dots to a time-varying Zeeman coupling
$(g\mu_B {\bf S}\cdot {\bf B})(t)$, which is controlled through
the local magnetic field ${\bf B}$ or the local $g$-factor in
semiconductor nanostructures. Localized magnetic fields can be
generated with the magnetic tip of a scanning force microscope.
The local $g$-factor can be modified by external bias voltage. An
effective Zeeman field may also be realized by exchange spin
coupling to ferromagnetic dots \cite{PRA1998}, but spin exchange
couplings for single spin rotations are not our choice for
implementing free spin quantum computation.

Considering the problem relating to decoherence in semiconductor
nanostructures, we prefer to use another method for the
manipulation of single electron spins, i.e., the fast controls of
single spin coherence using ultrafast optical pulses. It has been
experimentally demonstrated that optical tipping pulses with a
frequency below the band gap of the semiconductor nanostructures
can create an effective magnetic field on the order of 20 T via
the optical Stark effect, which can induce substantial rotations
of electron spins at femtosecond scales\cite{Gupta}. Meanwhile,
spin-flip Raman transitions using the adiabatic process of two
ultrafast laser pulses\cite{adiabatic} can also fully control
single spin rotations in semiconductor quantum dots at picosecond
or femtosecond scales\cite{PRL1,Chen04}. These optical controls of
single spin rotations are technically attractive for practical
manipulation. As it has been demonstrated experimentally that the
typical decoherence time of a electron spin in semiconductor
nanostructures is about 50 $\mu s$, the bias voltage pulse for
single electron tunnelings has the time scale of a few 100 ps.
Fig.~4 tells us that the single spin rotations must be completed
much faster than the bias voltage pulses. Thus, the ultrafast
optical pulses for manipulating single spin at picosecond to
femtosecond scales are required to reduce decoherence effects.

At last, we shall also discuss the initialization and read-out of
single electron spins in this scheme. A static uniform magnetic
field can be applied to split the spin up state $|0 \rangle$ and
the spin down state $|1 \rangle$ by the Zeeman energy for
initialization. A single-shot read-out of the electron spin states
in qubit dot can be realized by coupling the unit cell to a
quantum point contact (QPC). Explicitly, one can tune a bias
voltage pulse, e.g. $V_i^{LR}$, to lower the on-site energy of
dots $C_i, D_i$ such that the electron will remain in the qubit
dot if it is in the state $|0\rangle$, otherwise it will tunnel to
the dots $C_i, D_i$ and then return back to the qubit dot after
the pulse if it is in the state $|1\rangle$. By measuring the
charge current through the QPC channel, $I_{QPC}$, one can detect
changes in charge that result from the electron tunneling between
the qubit dot and ancilla dots. In this way, we can measure the
single electron spin states in qubit dots through the QPC as a
charge detector. Such a measurement has been experimentally
realized in quantum dots\cite{spinm}.

In summary, combining the spin two-qubit CNOT gate with single
spin rotations, a universal quantum computation can be achieved
without using spin-spin interactions. Implementing free fermion
quantum computation is a very challenge subject in
principle\cite{no-go1,no-go2}. Here we are able to achieve such an
implementation relying basically on charge-to-spin conversion
through QCA. QCA offers an intrinsic charge coupling of two
electrons, which is more effective than the use of beam splitters
plus charge detectors\cite{free}. Since spin exchange interaction
is much weaker than electron Coulomb interaction (by the order of
$10^{-3}$), such an implementation of free spin quantum
computation has the advantage of being robust against the
technical difficulties of generating strong spin-spin
interactions. We hope that the realization of this scheme will
bring a new challenge to semiconductor spintronics and the
development of nanotechnology.

\begin{acknowledgments}
One of the authors (WMZ) would like thank Profs. G.J.~Milburn and
L.J.~Sham for useful discussions. This work is supported by the
National Science Council of Republic of China under Contract
No.~NSC-93-2120-M-006-005, No.~NSC-93-2112-M-006-011,
No.~NSC-93-2112-M-006-019 and National Center for Theoretical
Science, Taiwan.
\end{acknowledgments}


\end{document}